\documentclass{JAC2000}
\usepackage{graphicx}

\newcommand{\iso}[2]{\textsuperscript{#1}#2}
\newcommand{\ion}[2]{#1\textsuperscript{#2+}}
\newcommand{\aion}[3]{\textsuperscript{#1}#2\textsuperscript{#3+}}
\newcommand{\mion}[3]%
{\textsuperscript{#1}#2$_{\mathrm{#3}}$\textsuperscript{+}}

\begin{document}
\title{%
DESIGN OF THE 7 MEV/U, 217 MHZ INJECTOR LINAC FOR THE PROPOSED 
ION BEAM FACILITY FOR CANCER THERAPY\\ AT THE CLINIC IN HEIDELBERG}

\author{%
B. Schlitt, GSI, Planckstra\ss e 1, D-64291 Darmstadt, Germany\\
A. Bechtold, U. Ratzinger, A. Schempp, IAP, Frankfurt am Main, Germany}

\maketitle

\begin{abstract} 
A dedicated clinical synchrotron facility for cancer therapy using
energetic proton and ion beams (C, He and O) has been designed
at GSI for the Radiologische Universit\"atsklinik at Heidelberg, Germany. 
The design of the injector linac is presented.
Suitable ion sources are discussed 
and results of ion source test measurements are reported.
The LEBT allows for switching between two ion sources.
A short RFQ accelerates the ions from 8 keV/u to 400 keV/u. 
It is followed by a very compact beam matching section and a
3.8 m long IH-type drift tube linac for the acceleration to 7 MeV/u. 
Both rf structures
are designed for a resonance frequency of 216.816 MHz and 
for ion mass-to-charge ratios $A/q \le 3$ 
(\aion{12}{C}{4}, \mion{}{H}{3}, \aion{3}{He}{}, \aion{16}{O}{6}).
\end{abstract}

\section{INTRODUCTION}

\begin{table}[b]
\begin{center}
\caption{%
Major parameters of the injector linac.
}
\label{t:para}
\begin{tabular}{lr @{\hspace{2mm}} l}
\\[-3mm]
\hline\\[-3.5mm]
Design ion                    & \aion{12}{C}{4} & \\
Operating frequency           & 216.816 & MHz \\
Final beam energy             & 7 & MeV/u \\
Pulse currents after stripper & $\approx 100$ & e$\mu$A \ion{C}{6} \\
                              & $\approx 0.7$ & mA protons \\
Beam pulse length             & $\le 200$ & $\mu$s @ $\le 5$ Hz \\
Duty cycle                    & $\le$ 0.1 & \% \\
Norm.\ transverse exit & & \\
~~~beam emittances (95\,\%) \textsuperscript{1}
                              & $\approx 0.8$ & $\pi$ mm mrad \\
Exit momentum spread \textsuperscript{1}
                              & $\pm 0.15$ & \% \\
Total injector length \textsuperscript{2}
                              & $\approx$ 13 & m \\
\hline
\end{tabular}
\\[1mm]
\textsuperscript{1} 
{\small Not including emittance growth effects in the stripper foil.} \\
\textsuperscript{2} 
{\small Including the ion sources and up to the foil stripper.}
\end{center}
\end{table}

Since December 1997 nearly 70 patients have been 
treated successfully with energetic carbon ion beams
within the GSI cancer treatment program.
Advanced technologies like the intensity-controlled rasterscan method
for 3-di\-men\-sion\-ally conformal tumor treatment
using pencil-like ion beams and an active control of the 
beam intensity, energy, position and width during the irradiation
have been developed \cite{kraft98,eik98}.
The developments and experiences of this program 
led to a proposal for a hospital-based ion
accelerator facility for the clinic in Heidelberg
\cite{eik00}.
It consists of a 7 MeV/u injector linac 
and a 6.5 Tm synchrotron \cite{doli00}
to accelerate the ions to final energies of 50 to 430 MeV/u.
Three treatment areas 
(two isocentric ion gantries and one fixed horizontal beam line) 
are proposed to treat about 1000 patients/year.
To cover the specific medical requirements,
the accelerator facility is designed to deliver both beams of low-LET 
(linear energy transfer) ions (p, He) and high-LET ions (C, O).
The requested maximum beam intensities at the irradiation point
are $1 \times 10^{9}$ carbon ions/spill 
and $4 \times 10^{10}$ protons/spill.
Only active and no passive beam delivery systems are planned.

\section{INJECTOR LAYOUT}

\begin{figure*}[bht]
\centering
\includegraphics*[width=0.81\textwidth]{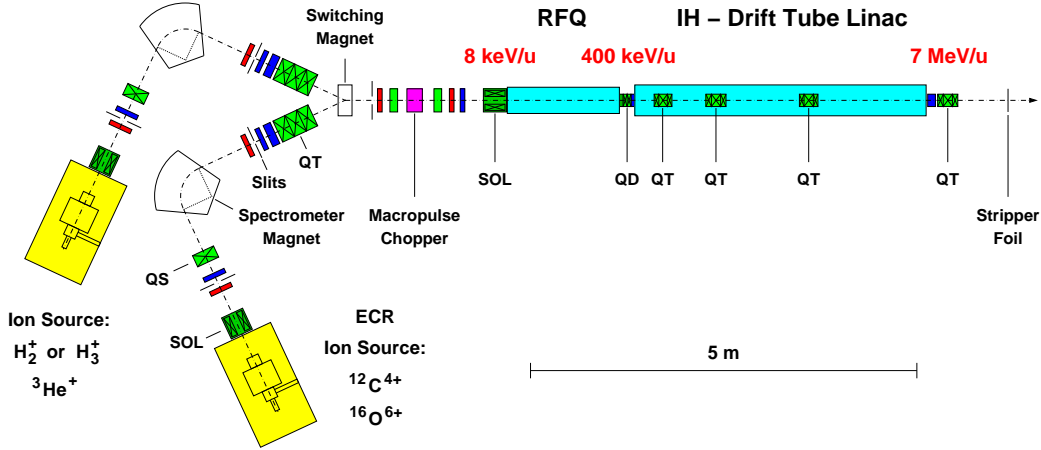}
\caption{%
Schematic drawing of the injector linac.
SOL $\equiv$ solenoid magnet,
QS, QD, QT  $\equiv$ magnetic quadrupole singlet, doublet, triplet.
}
\label{linac}
\end{figure*}

A compact injector linac with a total length of about 13~m 
has been designed (Fig.~\ref{linac} and Table \ref{t:para}).
To provide a fast switching between low and high-LET ion beams,
the analyzed beams from two ion sources running in parallel 
can be selected by a switching magnet before injection into 
the rf linac.
For the production of the high-LET ion beams
an ECR ion source (ECRIS) is proposed.
For the production of the low-LET ion beams
the installation either of an ECRIS of the same type 
or of a much more compact, cheaper and simpler
gas discharge ion source is discussed.

To form short beam pulses,
a fast macropulse chopper will be used in the common 
straight section of the LEBT line.
For the intensity-controlled rasterscan method
different beam intensities within an intensity range of 1/1000
are requested for each individual synchrotron cycle.
The required controlled beam intensity variation 
will be performed already along the LEBT line by
changing the driving currents of the quadrupole triplet magnets
following each spectrometer section from pulse to pulse.

The 21 MV rf linac \cite{schlitt98}
is designed for ion mass-to-charge ratios $A/q \le 3$ 
and an operating frequency of 216.816 MHz.
It has a total length of only about 5.5 m
and consists of two cavities ---
a short RFQ structure and an efficient IH-type drift tube linac.
For stripping off the remaining electrons prior to
injection of the ions into the synchrotron,
a thin foil stripper located about 1 m behind of the DTL
is used for all ion species.

To avoid contaminations of the helium ion beams with ions from other
elements having the same $A/q$,
the use of \iso{3}{He} instead of \iso{4}{He} is proposed.
To reduce space-charge effects along the complete injector
linac in case of hydrogen ion beams 
and to increase the extraction voltage of the ion source,
the production and acceleration of molecular 
\mion{}{H}{2} or \mion{}{H}{3} ion beams is planned.
The molecules are breaking up into protons at the stripper foil.

\section{ION SOURCES}

To achieve the demanded beam intensities at the irradiation
point with only moderate requirements for the ion sources,
a multiturn-injection procedure with an accumulation factor of 10
is proposed for the synchrotron.
Considering reasonable loss factors for the complete 
accelerator chain and the beam lines,
the ion currents required from the ion sources
range from roughly 100 e$\mu$A \ion{O}{6} 
to about 650 $\mu$A \mion{}{H}{2} (Table \ref{t:source-req}).

\begin{table}[b]
\begin{center}
\caption{%
Ion currents $I_{\mathrm{ion}}$ required from the ion sources
and ion source potentials $V_{\mathrm{IS}}$ needed for a beam energy
in the LEBT of 8 keV/u.}
\label{t:source-req}
\begin{tabular}{cccc}
\\[-2.5mm]
\hline\\[-4mm]
Ion species & $I_{\mathrm{ion}}$ & $V_{\mathrm{IS}}$ & Ion species \\
from source & [$\mu$A] & [kV] & to synchrotron \\
\hline\\[-3.5mm]
\aion{16}{O}{6}  &  100 & 21.3 & \aion{16}{O}{8} \\
\aion{12}{C}{4}  &  130 & 24   & \aion{12}{C}{6} \\
\aion{3}{He}{1}  &  320 & 24   & \aion{3}{He}{2} \\
\mion{1}{H}{2}   &  650 & 16   & protons \\
\mion{1}{H}{3}   &  440 & 24   & protons \\
\hline
\end{tabular}
\end{center}
\end{table}

\subsection{ECR ion source}

A high-performance 14.5 GHz fully permanent magnet ECRIS
called SUPERNANOGAN has been developed at GANIL \cite{sortais98}
and is commercially available from PANTECHNIK S.A., France.
To check the suitability of the source for the therapy injector,
test measurements have been performed at the ECRIS test bench
at GANIL\@.
The required ion currents could be exceeded by at least 50\,\% 
(\ion{C}{4}, \mion{}{H}{2})
up to a factor of about 3 (\ion{O}{6}, \ion{He}{1})
in a stable DC operating mode.
The rf power transmitted by the rf generator was
about 100 W for the extraction of 1.1 mA \ion{He}{1}
up to about 420 W for a 200 e$\mu$A \ion{C}{4} beam.
The measured normalized 90\,\% transverse beam emittances
are $< 0.5 \; \pi$ mm mrad for 280 e$\mu$A \ion{O}{6},
about 0.6 -- 0.65 $\pi$ mm mrad for \ion{C}{4} and \ion{He}{1+,2}
and roughly 0.7 $\pi$ mm mrad for a 1.5 mA proton beam.
In the latter case, the measured values may be limited by the 
acceptance of the spectrometer system at GANIL\@.
During the tests,
some high-volt\-age problems occurred above an extraction voltage
of about 20 kV\@.
Meanwhile, these problems have been analyzed by PANTECHNIK and
some improvements will be tested soon.
The solution of these problems is essential for the therapy injector
since extraction voltages of up to 24~kV are required for 
a beam energy of 8 keV/u (Table \ref{t:source-req}).
However,
alternative high-performance ECR ion sources using electromagnets
are available,
which can be operated at the required source potentials.
For instance, the ECR4-M type ECRIS available from PANTECHNIK
or the 10 GHz NIRS-ECR operated at HIMAC\@.

\subsection{Gas discharge ion source}

Besides economical reasons,
a gas discharge ion source optimized for the production 
of singly charged ions has several advantages.
In contrast to ECR ion sources,
where the \mion{}{H}{3} fraction of the extracted hydrogen ion beams
is only a few percent,
it can be optimized to more than 90\,\% in case of 
a gas discharge ion source at low arc currents 
\cite{holl00,holl-diss,holl-priv}.
The acceleration of \mion{}{H}{3} ion beams
has the important advantage that the rf power levels in the
linac cavities can be identical in case of \aion{12}{C}{4}
and for hydrogen ion beams.
Hence, faster switching between both beams would be possible 
as well as a very stable and reliable operation of the cavities.
Furthermore, 
higher ion currents can be extracted easily
and very high beam qualities are achieved.
For example,
for a 9 mA \ion{He}{} beam extracted with an extraction voltage of 17 kV 
a normalized 80\,\% transverse beam emittance of 0.003 $\pi$ mm mrad
was measured using the high-current high-brilliance gas discharge
ion source developed at the 
Institut f\"ur Angewandte Physik (IAP) at the University of Frankfurt
\cite{volk94}.
With the same source,
current densities of more than 40~mA/cm\textsuperscript{2}
could be achieved easily for \mion{}{H}{3} beams
with \mion{}{H}{3} fractions of about 94\,\% \cite{holl-diss,holl-priv}.

\section{RF LINAC}

A compact four-rod like RFQ structure for the acceleration from 
8 keV/u to 400 keV/u has been designed at the IAP\@.
The electrode length is 1.35 m, the electrode voltage is 70~kV
and the expected rf peak power is about 100~kW
at a low duty cycle around 0.1\,\%.
For matching the output beam parameters to the values required
at injection into the IH-DTL a very compact scheme is proposed.
For bunching the beam in the longitudinal phase plane
a drift tube directly following the RFQ structure will be integrated
into the RFQ tank.
Results of PARMTEQ simulations of the RFQ
as well as first results of model measurements and MAFIA simulations
regarding the integration of the drift tube are reported in
Ref.\ \cite{bechtold00}.

For focusing the beam in both transverse phase planes 
and for correction of small angular deviations of the beam 
at the RFQ exit,
a magnet unit consisting of an xy-steerer and a 
magnetic quadrupole doublet is flanged to the RFQ tank.
The unit has a total length of only 15 cm.
It is followed by a diagnostic chamber of 5 cm length,
which contains a capacitive phase probe and a beam transformer.
The simulation of the particle dynamics along the
matching section is included in the simulations of the
RFQ and of the IH-DTL \cite{schlitt98,bechtold00}.

The IH-type drift tube linac for the acceleration from 
0.4 MeV/u to 7 MeV/u consists of four KONUS \cite{rati98} sections
housed in the same cavity of about 3.8 m in length and 
30 cm in diameter \cite{schlitt98,minaev99}.
It consists of 56 accelerating gaps and three integrated magnetic
quadrupole triplet lenses.
The expected rf peak power is about 1 MW\@.
To achieve an approximately constant maximum on-axis electric field
of about 18 MV/m along the whole structure,
the gap voltage distribution is ramped from about 200 kV at the
low-energy end to about 480 kV at maximum.
By careful optimization of the individual KONUS sections,
the acceptance of the structure was increased
to about 1.3 $\pi$ mm mrad (norm.) in the transverse phase planes,
and to about 3.0 $\pi$ ns keV/u in the longitudinal plane.

\begin{figure}[t]
\centering
\includegraphics*[width=82.5mm]{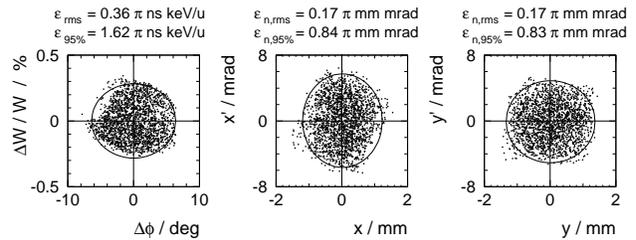}
\caption{%
Particle distributions at the stripper foil.
The ellipses contain 95\,\% of the particles.
}
\label{emi-out}
\end{figure}

Behind the IH-DTL the beam is focused on the stripper foil by another
magnetic quadrupole triplet.
The particle distributions at the stripper foil resulting
from particle tracking simulations along the DTL using the LORASR code
are presented in Fig.~\ref{emi-out}.
The particle distributions used at injection into the DTL
have been matched to the results of the RFQ simulations.
The 95\,\% emittance areas at DTL injection
are 1.3 $\pi$ ns keV/u in the longitudinal phase plane and
0.7 $\pi$ mm mrad (norm.) in both transverse planes.
The transverse beam emittances are based on the values measured
for the ECR ion sources under discussion.
The relative growth of the 95\,\% ellipse areas along the DTL
is about 22\,\% in all three phase space projections,
the rms emittance growth amounts to about 10\,\% in each plane.
Beam envelopes along the DTL have been presented already in earlier
publications \cite{schlitt98,minaev99}.
The current limit for the IH-DTL for ions with $A/q = 3$ 
resulting from beam dynamics simulations
is larger than 20 e$\mu$A \cite{minaev99}.
The momentum spread of the ion beam at the stripper foil is
about $\pm 0.15$\,\%.
It will be increased due to energy-straggling effects in the foil.
To enhance the injection efficiency into the synchrotron,
the momentum spread will be reduced to $\le \pm 0.1$\,\%
by a debuncher cavity installed in the synchrotron injection beam line.
An 1:2 scaled rf model of the IH-DTL structure 
is designed at present at GSI\@.
First model measurements are scheduled for the second quarter of
the next year.

\section{ACKNOWLEDGEMENTS}

We would like to thank C.~Bieth, S.~Kantas, O.~Tasset and E.~Robert
(PANTECHNIK) for performing the SUPERNANOGAN test measurements
at GANIL\@.
The fruitful cooperation of L.~Dahl (GSI) in the LEBT design
is greatly acknowledged.

\end{document}